\documentclass[final,5p,times,twocolumn]{elsarticle} 

\usepackage{lineno}

\usepackage{graphicx}

\usepackage{amssymb}

\usepackage{amsmath}

\usepackage{blindtext}
\usepackage{url}

\usepackage{placeins}

\journal{Nuclear Instruments and Methods A}

\begin{document}

\begin{frontmatter}



\title{Fast Frontend Electronics for high luminosity particle detectors}




\author[1,2]{M.~Cardinali\corref{cor1}} \ead{cardinal@kph.uni-mainz.de}
\author[1]{O.~Corell}
\author[1]{M.I.~Ferretti Bondy}
\author[1]{M.~Hoek}
\author[1]{W.~Lauth}
\author[1]{C.~Rosner} 
\author[1]{C.~Sfienti}
\author[1]{M.~Thiel}
\address[1]{Institut f\"{u}r Kernphysik, Johannes Gutenberg-University Mainz, Mainz, Germany}
\address[2]{Helmholtz Institut Mainz, Mainz, Germany}

\cortext[cor1]{Corresponding author}

\begin{abstract}

Future experiments of nuclear and particle physics are moving towards the high luminosity regime, in order to access suppressed processes like rare B decays or exotic charmonium resonances. In this scenario, high rate capability is a key requirement for electronics instrumentation, together with excellent timing resolution for precise event reconstruction. The development of dedicated FrontEnd Electronics (FEE) for detectors has become increasingly challenging. A current trend in R\&D is towards multipurpose FEE which can be easily adapted to a great variety of detectors, without impairing the required high performance.\\
We report on high-precision timing solutions which utilise high-bandwidth preamplifiers and fast discriminators providing Time-over-Threshold information, which can be used for charge measurements or walk corrections thus improving the obtainable timing resolution. The output signal are LVDS and can be directly fed into a multi-hit TDC readout. 

The performance of the electronics was investigated for single photon signals, typical for imaging Cherenkov detectors. The opposite condition of light signals arising from plastic scintillators, was also studied. 
High counting rates per channel of several~MHz were achieved, and a timing resolution of better than 100~ps could be obtained in a test experiment using the full readout chain.

\end{abstract}


\end{frontmatter}



\section{Introduction}

High luminosity experiments will play a key role for the future of particle and nuclear physics. In consequence, the next generation of detectors will feature increased complexity and require outstanding capabilities, e.g. high channel density setups will cope with extreme count rates. Nowadays the development of custom-designed FEE for specific applications has become very demanding in terms of costs and resources. The development of dedicated Application-Specific Integrated Circuits (ASICs), for example, has seen a dramatic increase in complexity and functionality~\cite{FEEASIC}. These reasons favor a multi-purpose approach to the development of new FEE.\\
The proposed FEE was developed to meet the challenging requirements of an imaging Cherenkov counter, the Barrel DIRC (Detection of Internally Reflected Cherenkov light)~\cite{Matthias} of the PANDA experiment~\cite{phy}. The detection of single photons will be of fundamental importance for this detector. In particular, the photodetector readout based on Microchannel Plate Photomultipliers (MCP-PMTs) will deliver fast ($\sim2$~ns) and small ($<5$~mV) signals, with an intrinsic resolution better than 40~ps after time walk correction~\cite{Albert}. High precision timing on Single Photo-Electron (SPE) ($\sigma_{time}<100$~ps) is needed for pattern reconstruction and background suppression. Moreover, the continuos $\bar{p}$ beam with a planned 20~MHz average interaction rate (with peaks up to 50~MHz), sets a minimum requirement of 100~kHz per channel. Last but not the least, the PANDA DAQ scheme imposes the use of triggerless FEE.\\
The third version of the TDC Readout Board (TRB3), developed at GSI, has been chosen as Data Acquisition System (DAQ) ~\cite{cahit}. The TRB3 offers FPGA-based TDCs with multi-hit capability and up to 10~ps time precision on leading and trailing edge time measurements (20~ps on average). High hit rate (67~MHz) capability and a unique flexibility further suit the PANDA Barrel DIRC requirements.\\
A design based on fast discriminators meets the required specifications and minimizes the information to digitize. A coarse but fast measurement of the deposited charge can be obtained through Time-over-Threshold ($ToT$). A modular design with a preamplifier Add-on card allows the realization of versatile FEE. Different detectors can be readout, for example plastic scintillators which usually deliver large signals produced by the energy loss of charged particles. The highly segmented plastic scintillator array for the future Neutron Detector (NDet) of the A1 Collaboration~\cite{bormio_thiel,SFB} was chosen as test case. The NDet will be instrumented with MultiAnode PMTs (MAPMTs), with a total of about 8000 readout channels. The future physics programme imposes high rate requirements (several~MHz per channel). Under these conditions, a $ToT$ logic is a valid choice, also because of the large amount of data. Particle identification to distinguish neutrons and Minimum Ionizing Particles (MIP) is an additional asset. 


\section{Frontend electronics design}

\begin{figure}[!htb]
\centering
\includegraphics[width=0.7\linewidth]{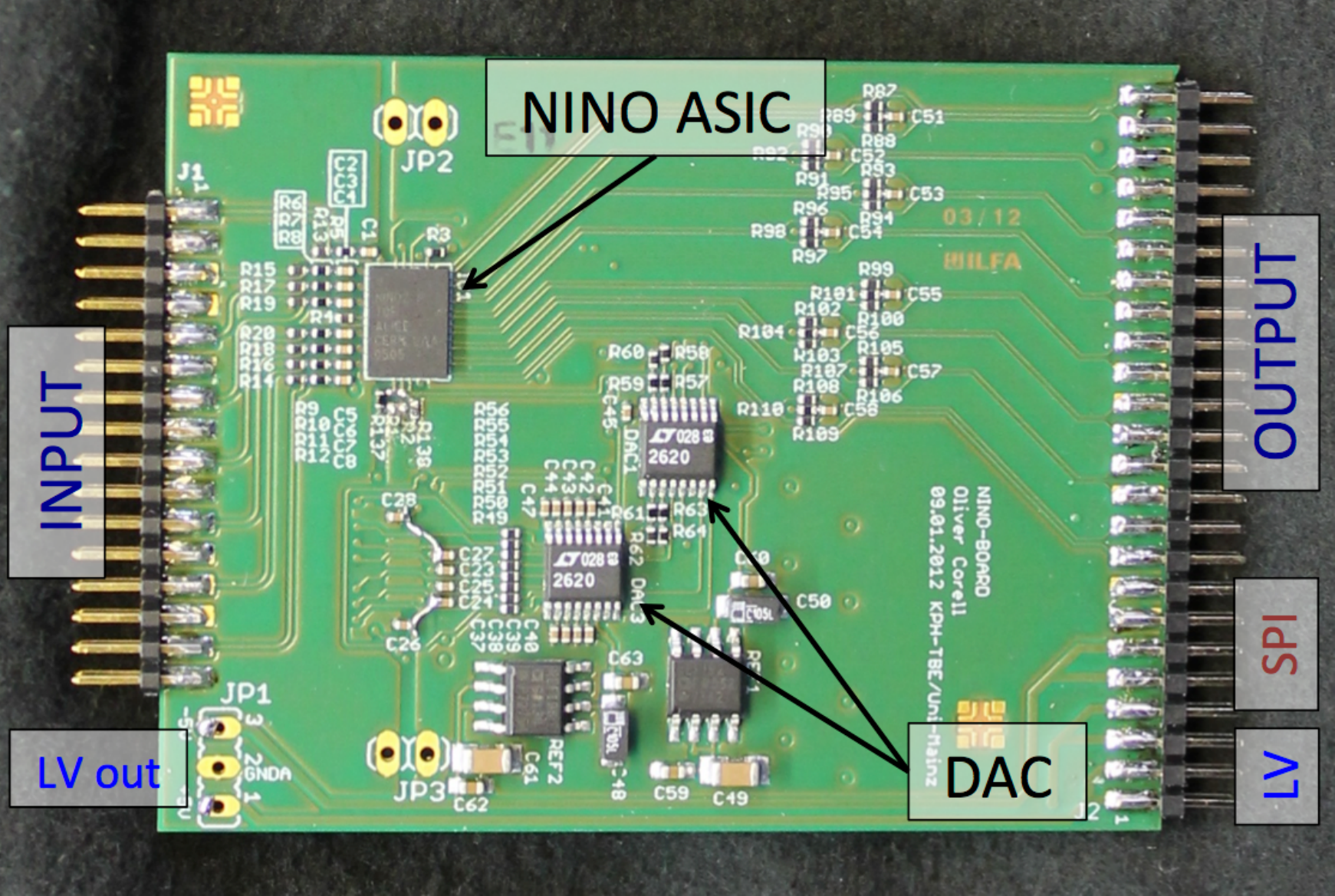}
\caption{Top view of the NINO discriminator PCB.}
\label{fee_discriminator}
\end{figure}

The development of this FEE was focused on the realization of a modular and versatile design. The main component is a discriminator card realized with a Printed Circuit Board (PCB) 8~cm long and 6~cm wide. The layout of the discriminator card was designed to be as compact as possible and to fit the MCP-PMT connectors. A picture of the discriminator card is shown in Fig.~\ref{fee_discriminator}. 
\\
The NINO ASIC, developed by the LAA project at CERN, was selected as core component for the discriminator card~\cite{nino}. The NINO chip features 8 channels and was chosen among other similar ASICs for its excellent timing performances, low noise and low power consumption~\cite{ninocern}. \\
The output of common PMTs, including MCP-PMTs, is usually single-ended, while the input of the NINO chip is differential. 
However the chip can work in ``single-ended mode''~\cite{despeisse}, where one channel input is connected to the PMT output while the other is left open. 
Two NINO chips per card (one per layer) are used for a total of 16 readout channels. Three commercial Digital to Analog Converter chips (DAC) set the working parameters of the NINO chips like common thresholds, hysteresis and pulse stretcher widths (offset for the input ToT). In particular, the DACs define voltage offsets (from 0 to 3.3V) which are added to the inputs of each NINO channel. These connections allow the definition of individual thresholds for every NINO channel, which by design only have one common threshold per chip. The communication with the DACs is via a Serial Peripheral Interface (SPI) connection. The NINO provides LVDS (Low Voltage Differential Signal) output, which is fed through a specific adapter to the TRB3 DAQ. The board requires an operating voltage of +5 V. The power consumption was measured to be about 70~mW/channel. Furthermore, the card provides the facility to power Add-on cards with $\pm5$~V.
\begin{figure}[htb]
\centering
\includegraphics[width=0.9\linewidth]{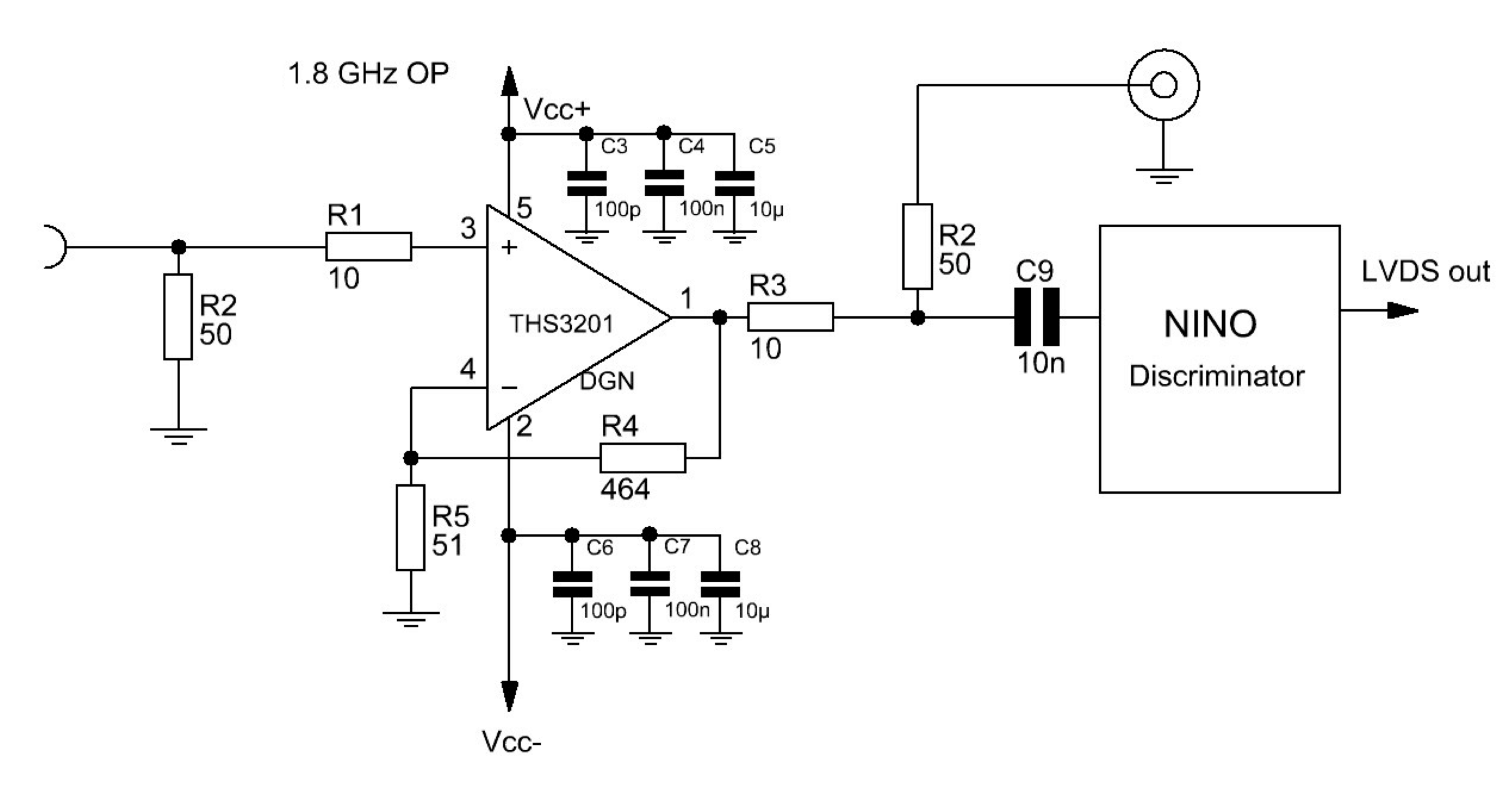}
\caption{Schematic of the preamplifier circuit for 1 channel.}
\label{fee_schematic}
\end{figure}

A preamplifier Add-on card was designed for the SPE scenario. The card, 4~cm long and 6~cm wide, has 16~ch (8~ch per layer). A current feedback amplifier chip\footnote{Texas Instruments, THS3201} provides about 20~dB gain, with 1.8 GHz unity-gain bandwidth. The schematic for one channel is shown in Fig.~\ref{fee_schematic}. The amplification shifts the input charge into a region ($> 200$~fC) where the intrinsic charge-dependent jitter of the NINO~\cite{nino} is small.


\FloatBarrier
\section{Frontend Electronics characterization}

The FEE cards were characterized under controlled conditions to understand the basic properties, as well as to investigate the behaviour of the single-ended scheme with individual thresholds. The discriminator and the preamplifier cards were first tested separately, using a programmable fast pulse generator\footnote{Tektronix AFG3252} (2~GS/s and 250~MHz bandwidth). The readout was realized with a fast oscilloscope\footnote{Tektronix DPO 5204} (10~GS/s and bandwidth of 2~GHz), so that the complete waveforms were available for offline analysis. Subsequently the FEE cards together with the TRB3 DAQ, were tested under realistic conditions using a fast laser setup.

\subsection{Discriminator card}
\label{sec_disc}

\begin{figure}[htb!]
\centering
\includegraphics[width=0.95\columnwidth]{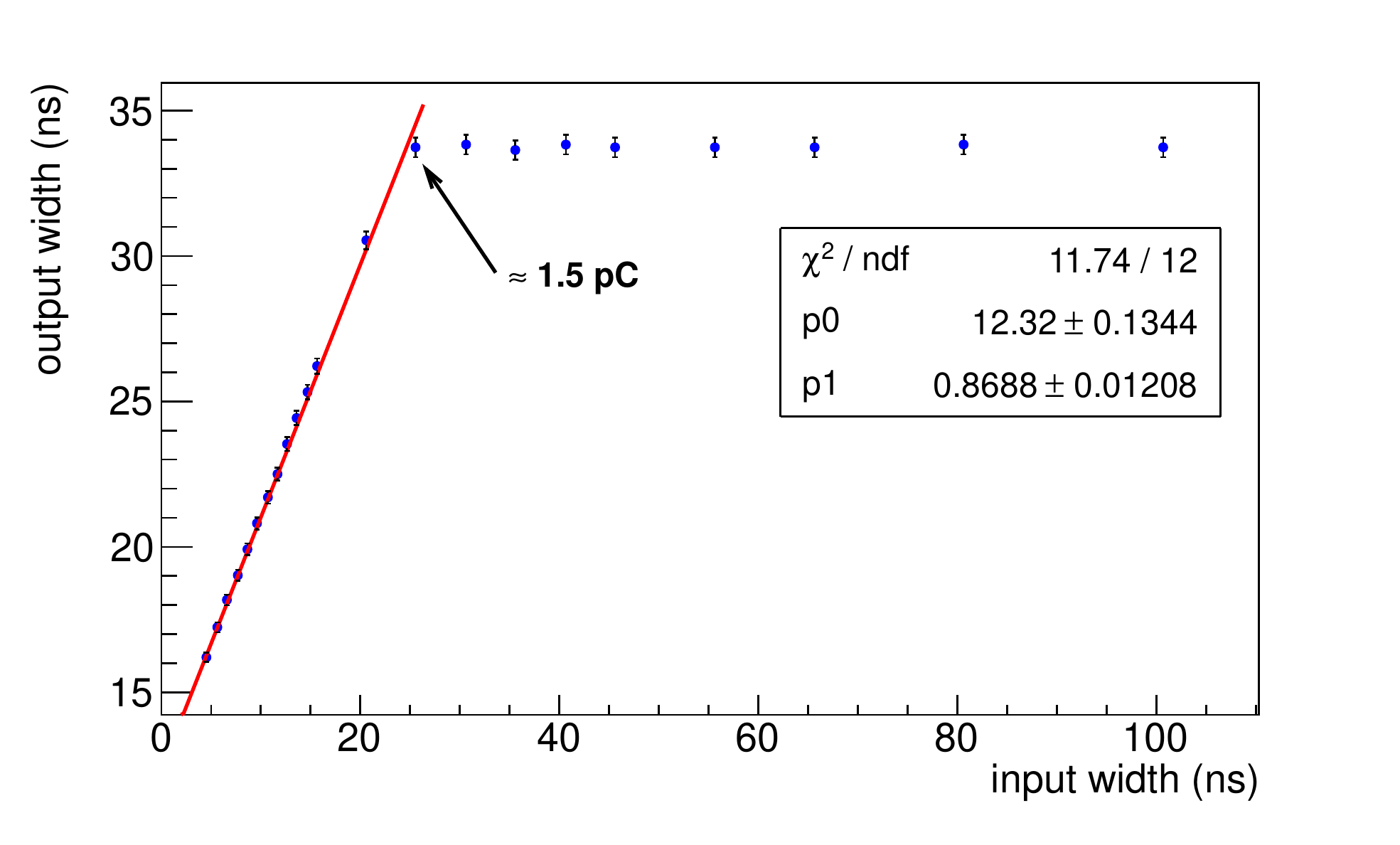}
\caption{Time-over-Threshold response of the discriminator card. The saturation of the card occurs at 1.5~pC, as indicated by the plateau.}
\label{fee_NINO_tot_response}
\end{figure}

The $ToT$ behaviour of the discriminator cards was analyzed choosing rectangular pulses with fixed amplitude and variable width as input. Fig.~\ref{fee_NINO_tot_response} shows the $ToT$ of the NINO output as function of the input width of the pulser signal. The response is perfectly linear up to almost 25~ns. At this value, corresponding to an input charge $q$ of 1.5~pC, the discriminator exhibits a saturation effect on the $ToT$. The result qualitatively agrees with the 2~pC limit on the input charge reported for the NINO ASIC~\cite{nino}. Observations showed no limits for $ToT>>100$~ns with $q<1.5$~pC. Further tests revealed that the leading edge time at saturation can be used up to an input charge of about 300~pC, while higher charge values introduce after-pulse effects after the main output signal.
 
\begin{figure}[htb!]
\centering
\includegraphics[width=0.95\columnwidth]{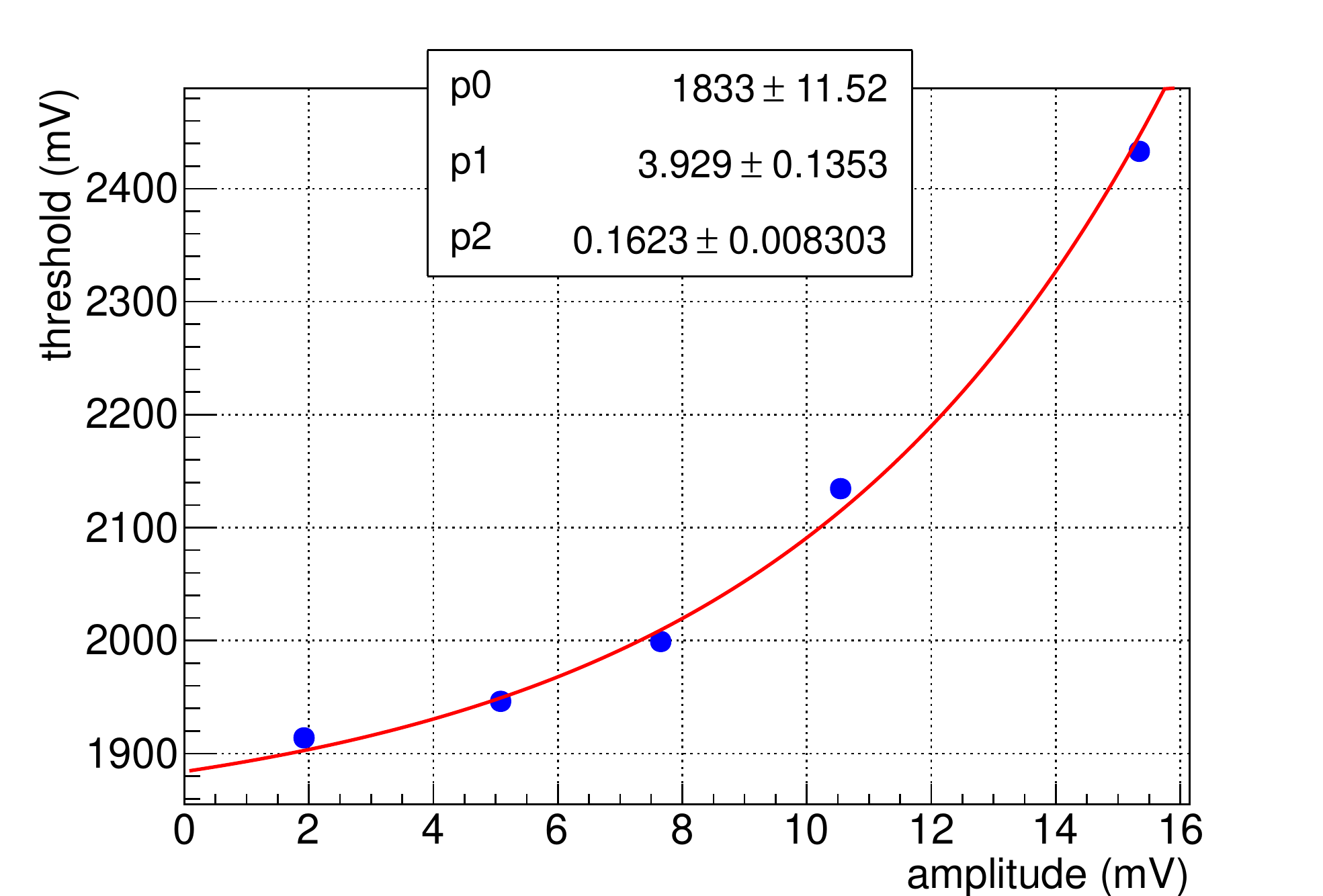}
\caption{Individual threshold calibration with short square pulser signals.}
\label{fee_thr_calibration}
\end{figure}

The individual thresholds realized in the discriminator PCB were calibrated using short square pulses ($ToT\approx 2~ns$). The average amplitude at which the NINO started producing output signals was recorded for various threshold settings. The result, shown in Fig.~\ref{fee_thr_calibration}, is clearly nonlinear and is fit with the following exponential function:
\begin{equation}
\label{eq_thr_cal}
V_{threshold} = p_{0} + e^{p_{1} + p_{2}\cdot amplitude}.
\end{equation}
The equation allows the precise setting of the threshold levels to the desired input amplitude.

\subsection{Preamplifier Add-on card}
The preamplifier cards are designed for use in a single photon environment. Consequently, the characterization with fast and small signals is of particular interest.
The  typical shape of the MCP-PMT response to single photons was reproduced with the pulser, using two gaussian distributions matching in the peak position. The standard deviation of the first gaussian represented the rising part of the signal and was set to $\sigma_{1}=0.6$~ns. The second distribution took into account the decay time of the signal, with a standard deviation set to $\sigma_{2}=1.2$~ns.

\begin{figure}[htb]
\centering
\includegraphics[width=0.95\columnwidth]{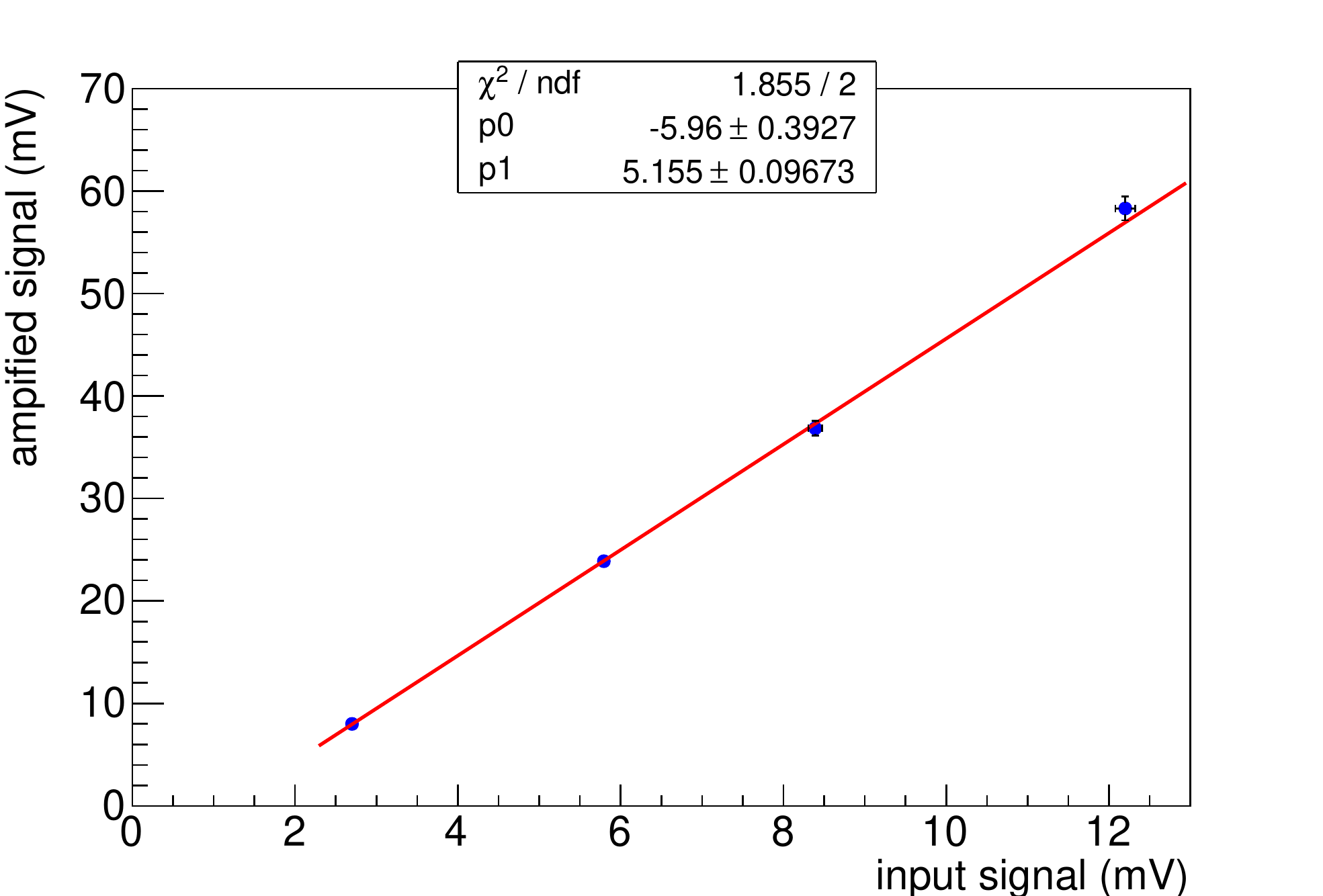}
\caption{Amplitude response of the preamplifier Add-on card to MCP-PMT-like pulser signals.}
\label{fee_preamp}
\end{figure}
The data in Fig.~\ref{fee_preamp} shows the amplitude response to different pulse heights. The trend is linear and returns a gain value $p1\approx5.2\pm0.1$. The result agrees with the design value of 20~dB considering the 50~$\Omega$ termination before the output (see Fig.~\ref{fee_schematic}).

\FloatBarrier
\subsection{Characterization with pulsed laser source}
The operation of the discriminator card together with the preamplifier was verified under realistic conditions. The most challenging scenario of single photon detection was reproduced by illuminating an MCP-PMT\footnote{PHOTONIS XP85012, used for PANDA~\cite{Albert}} with a fast laser pulser\footnote{PiLas laser diode, Advanced Laser Diode Systems A.L.S. GmbH, Germany} (35~ps FWHM, $\lambda\approx633$~nm) at a photon level of 0.3~$\gamma$/pulse. The photodetector was equipped with 4 FEE cards plus Add-on cards, for a total of 64 readout channels. The laser beam was focused on a central pixel and set to a trigger rate of 1~kHz. The setup was placed inside a light-tight box to shield the PMT from ambient light. Two TRB3 boards were used to readout the system.\\
\begin{figure}[htb!]
\centering
\includegraphics[width=0.95\columnwidth]{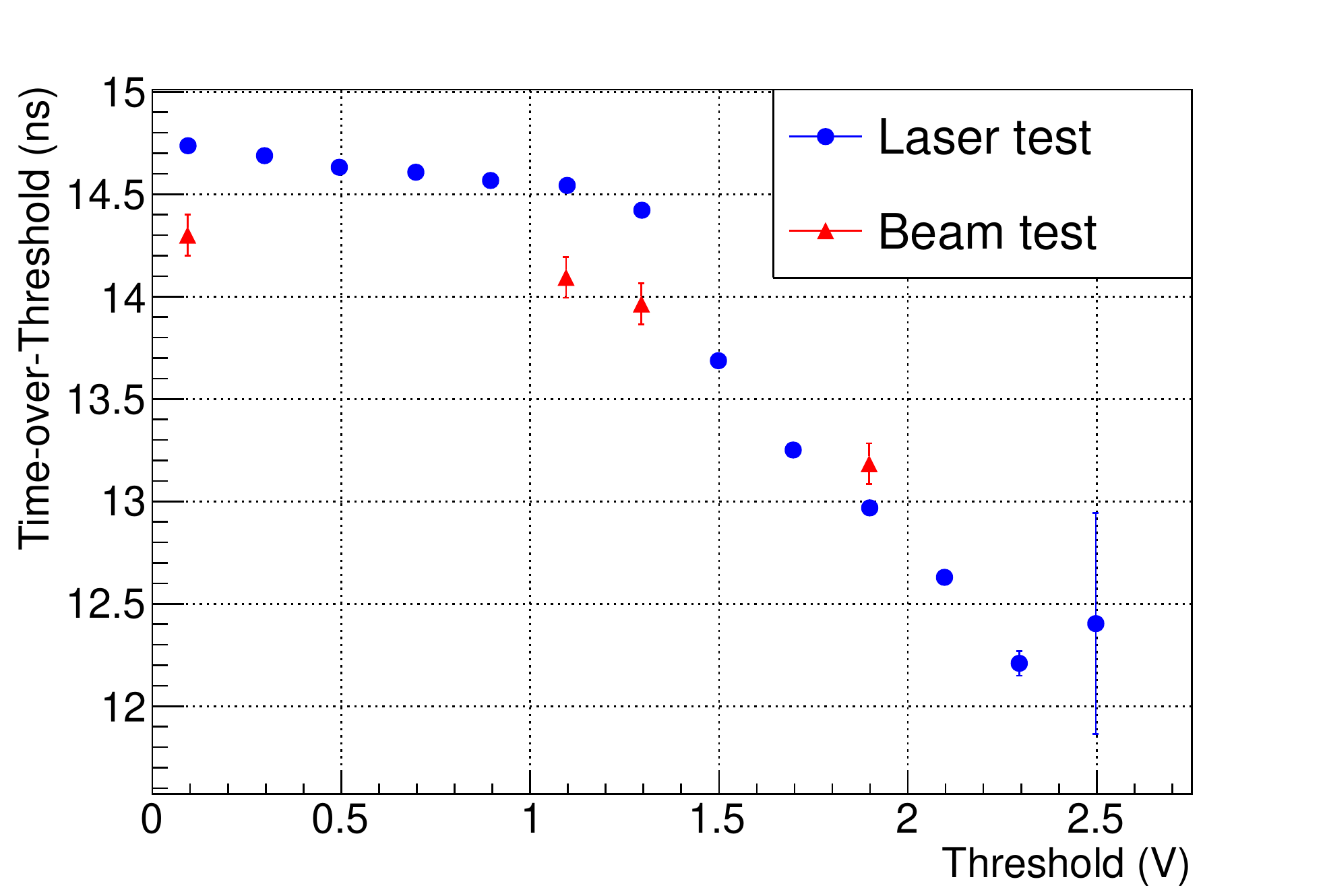}
\caption{Time-over-Threshold response to MCP-PMT signals as a function of the threshold (blue dots: laser test; red triangles: beam test)}
\label{fee_tot_behaviour}
\end{figure}

The behaviour of the ToT as function of the individual threshold (Fig.~\ref{fee_tot_behaviour}) allows the definition of a proper working threshold at the end of the plateau ($\approx1.3$ V), as well as serving as a cross check between laser and test experiment (see Section~\ref{sec_dirc}) data. 
\begin{figure}[htb!]
\centering
\includegraphics[width=0.95\columnwidth]{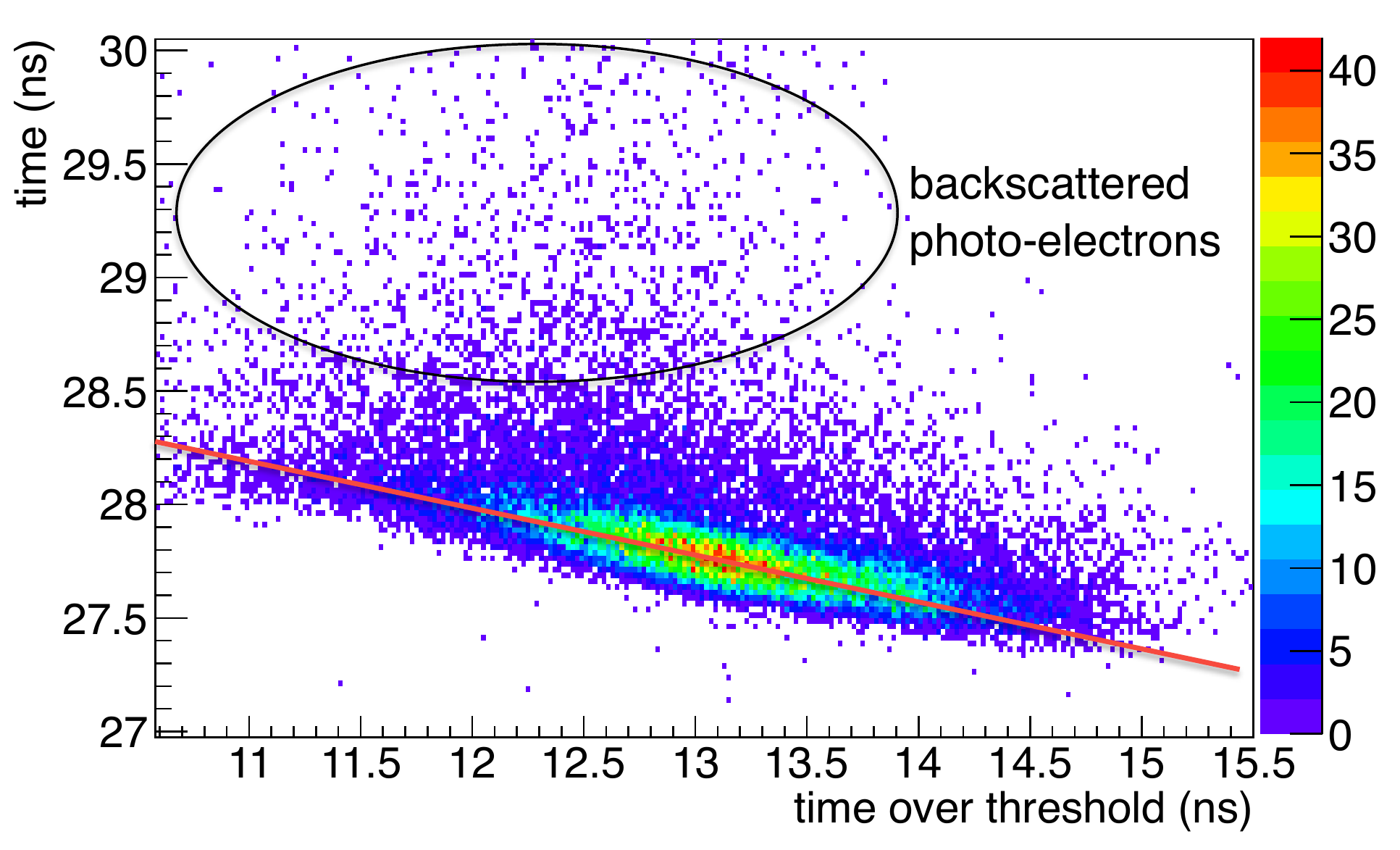}
\caption{Photon arrival time as function of Time-over-Threshold showing walk dependence for NINO. The red line shows the linear fit to the data.}
\label{tot_corr}
\end{figure}
The scatter plot in Fig.~\ref{tot_corr} shows the arrival time relative to the laser trigger $t_{meas}$ as function of the measured ToT. The signals at $t_{meas}>28.5$~ns belong to backscattered photo-electrons~\cite{Albert}. The correlation between ToT and $t_{meas}$ can be considered in first approximation linear and is used for a walk correction as given below: 
\begin{equation}
t_{corr}=t_{meas}-t^{*}\left(ToT\right)
\label{timewalk}
\end{equation}
where $t^{*}(ToT)= m\cdot ToT + q$ is the parametrisation of the distributions in Fig.~\ref{tot_corr}. \\
Defining the timing resolution $\sigma_{t_{corr}}$ as the standard deviation of $t_{corr}$, the walk correction returns $\sigma_{t_{corr}}\approx75$~ps, gaining an improvement of 40$\%$ with respect to the raw timing resolution ($\sigma_{t_{meas}}\approx122$~ps). This value represents the timing resolution for SPE of the full readout, and is already well below the 100~ps required by the PANDA Barrel DIRC.


\section{Test Experiment with an Imaging Cherenkov Counter}
\label{sec_dirc}
The full electronics chain comprising MCP-PMTs, FEE and TRB3 was extensively studied during a test experiment at the MAMI accelerator in Mainz~\cite{MAMI}. The continuous wave beam provides electrons with an energy of 855 MeV, a spot diameter of 1 mm and a negligible divergence~\cite{MAMIpol}. A small scale DIRC prototype with a fused silica radiator bar and an expansion volume filled with mineral oil was instrumented with a matrix of 4 MCP-PMTs, set to approximately $1\times10^{6}$ gain. 16 FEE cards (with Add-on cards) were used with a total of 256 channels read out with TRB3 boards (more details in~\cite{ioRICH,maria,Rosner}).
\begin{figure}[htb!]
\centering
\includegraphics[width=0.95\columnwidth]{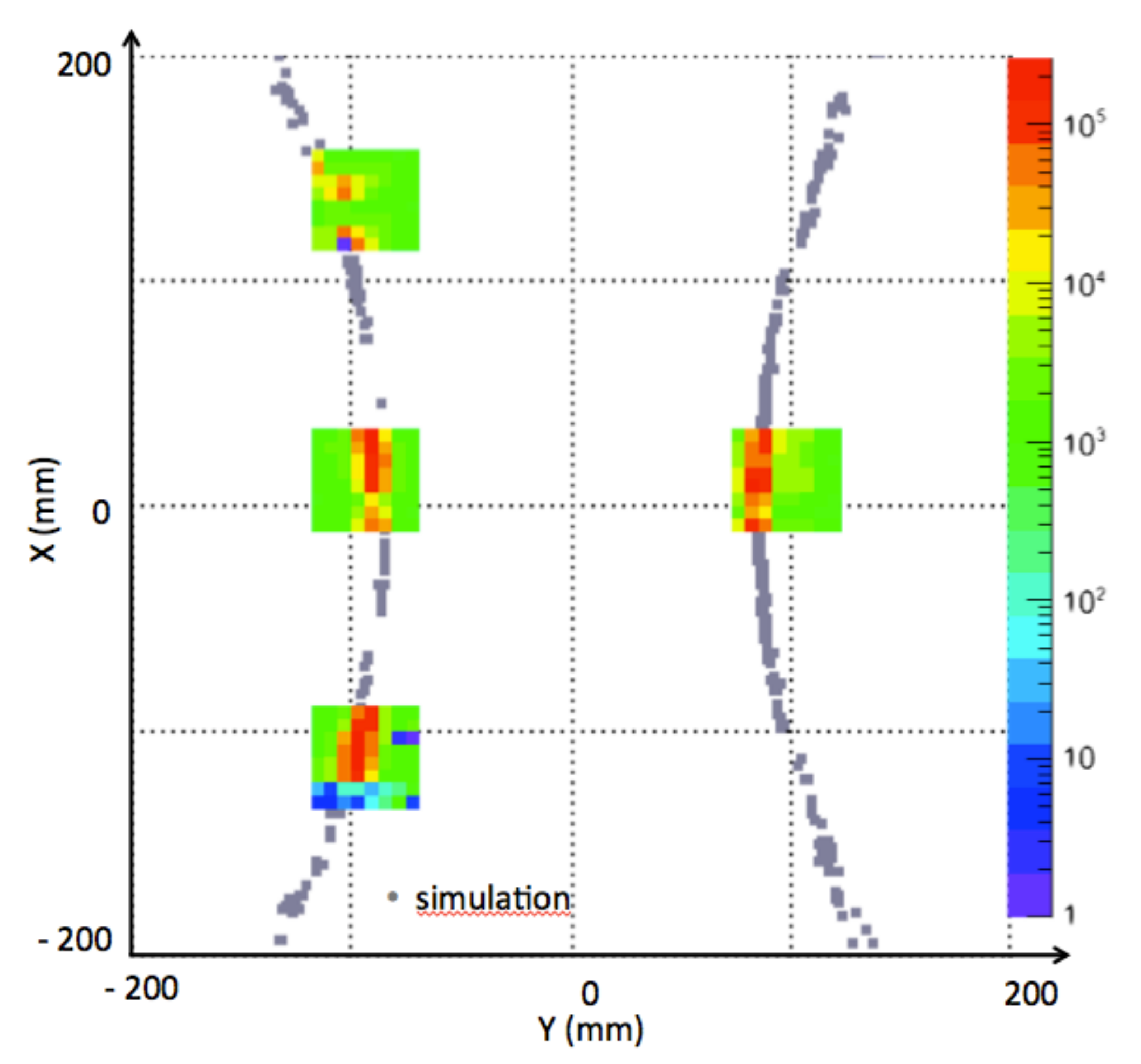} 
\caption{Observed Cherenkov pattern at $57^{\circ}$ incidence angle. Monte Carlo simulation is shown in grey.}
\label{poster}
\end{figure}

Fig.~\ref{poster} shows the result for a configuration for which the incidence angle between the radiator and the beam was $57^{\circ}$. The image shows an overlay of 1 million events. The Cherenkov pattern is clearly visible and was confirmed by varying the incidence angles. The results agree with a Monte Carlo simulation based on ray tracing~\cite{Carsten} (shown in grey and used only to verify the correct position). The FEE shows a good performance for single-photon detection under the anticipated working conditions.

\subsection{Timing studies}
\begin{figure}[htb!]
\centering
\includegraphics[width=0.95\columnwidth]{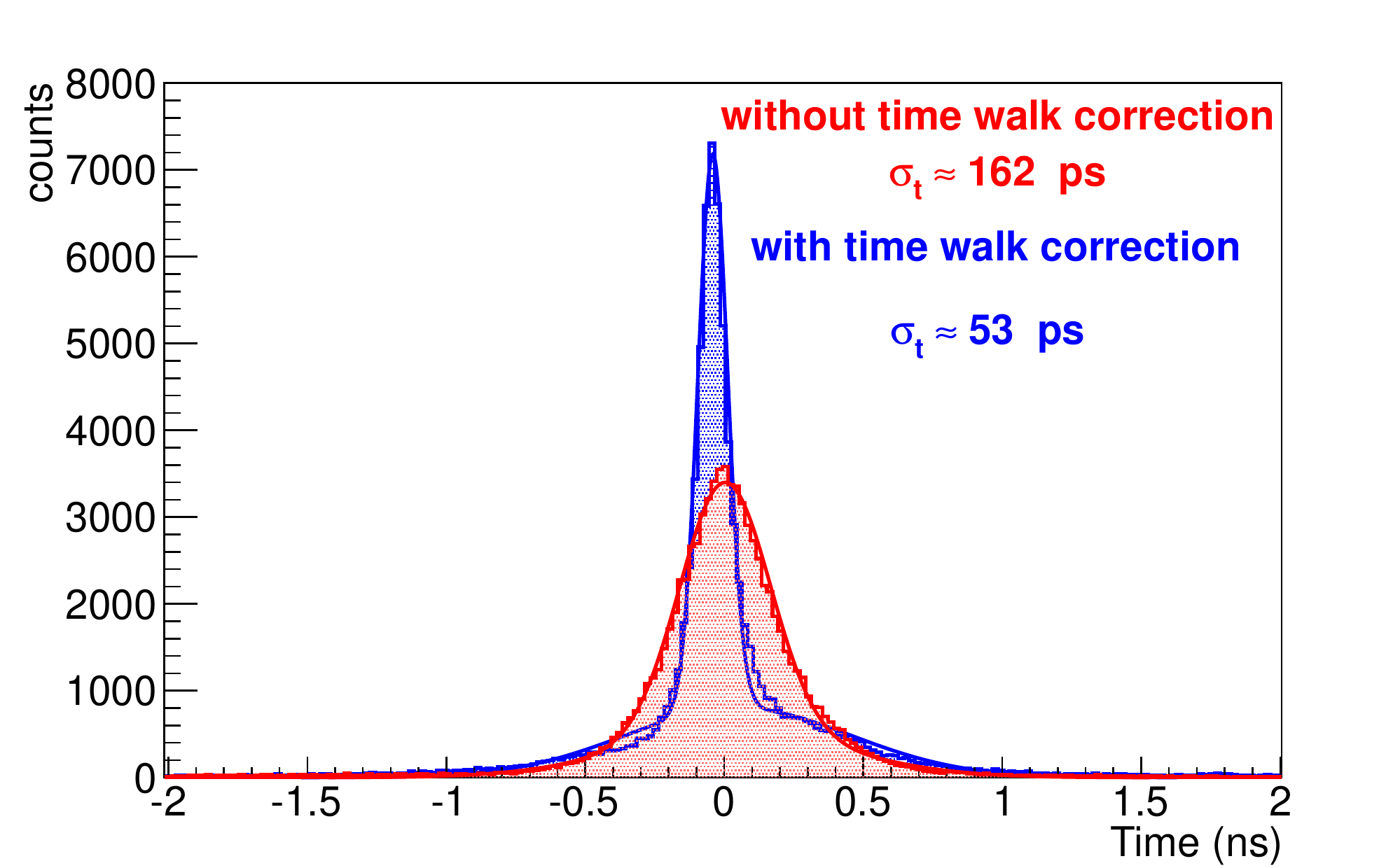} 
\caption{ Measured timing resolution between neighbouring MCP-PMT pixels: corrected data (blue), uncorrected data (red).}
\label{time_res_SAME_MCP}
\end{figure}

The FEE contribution $\sigma_{FEE}$ to the SPE timing resolution was measured taking advantage of charge sharing between neighbouring pixels in the same MCP-PMT. The low number of detected photons per event (2-3) allows the identification of charge sharing as SPE events. In this way the contribution of the MCP-PMT to the timing is minimised and the electronics determines the obtainable resolution. The results, shown in Fig.~\ref{time_res_SAME_MCP}, can be fit with a sum of two Gaussians taking into account the presence of back-scattered photo-electrons inside the MCP-PMT. The timing resolution $\sigma_{FEE}$ is defined as the standard deviation of the narrow peak. The walk correction improves $\sigma_{FEE}$ by about 70\%, with an excellent value of $\sigma_{SPE}\approx53$~ps.

The different running conditions (e.g. temperature and gain), between the laser characterization and the test experiment, require the optimisation for the coefficients of $t^{*}(ToT)$ in eq.~\ref{timewalk}. In particular the slope $m$ of the time walk parametrisation (see Fig.~\ref{tot_corr}) was tuned by varying its value found with the laser studies and looking for a minimum of the timing resolution $\sigma_{FEE}$. This simple algorithm further improves the timing resolution on average to $\sigma_{FEE}\approx40$~ps. The result is interpreted as the intrinsic resolution of the whole electronics chain.\\
The timing resolution of the prototype $\sigma_{prototype}$ was measured considering the time differences of pixels from different MCP-PMTs in the horizontal plane ($y\approx0$), which refers to photons with a symmetric path from the radiator to the focal plane.
\begin{figure}[htb!]
\centering
\includegraphics[width=0.95\columnwidth]{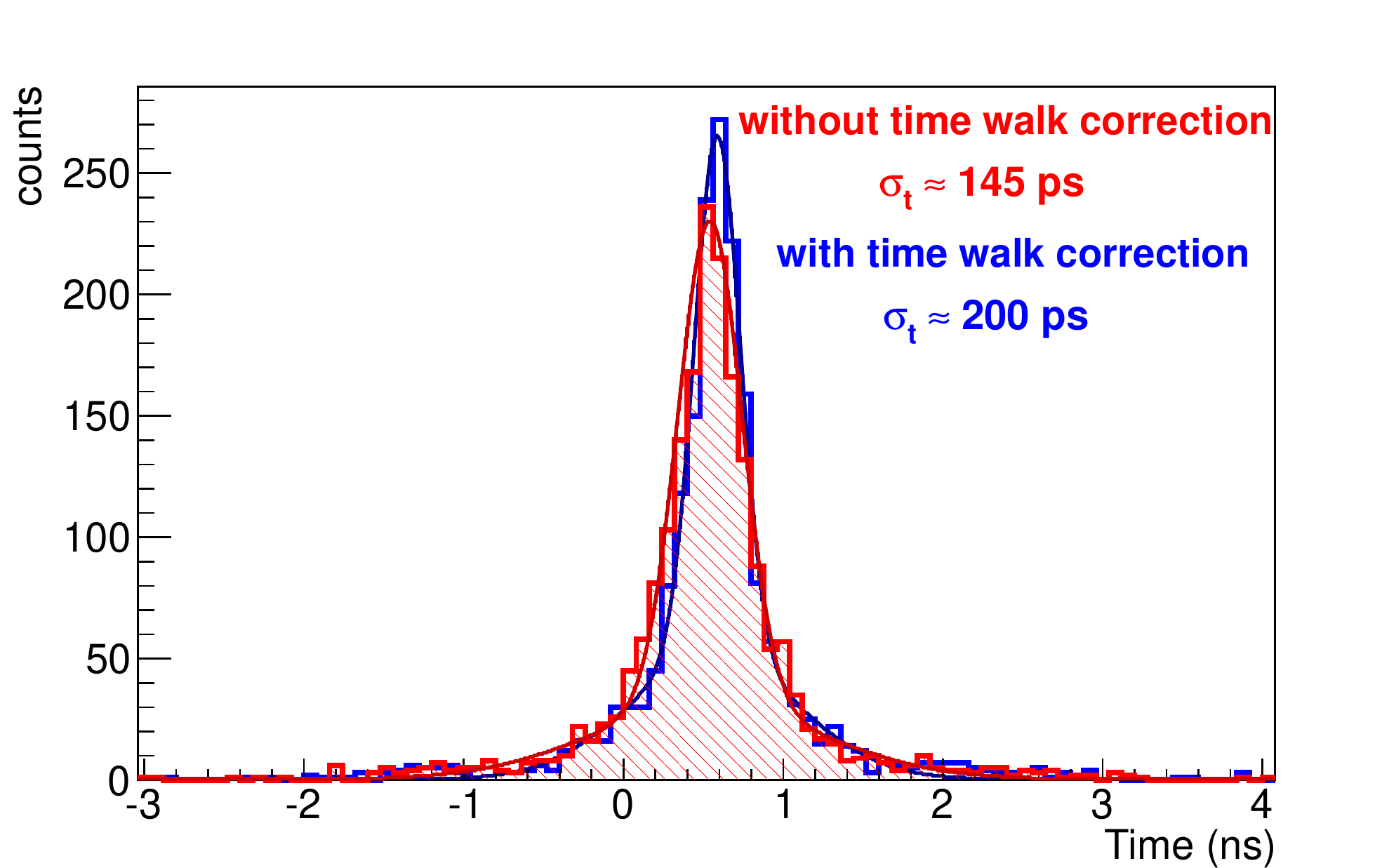} 
\caption{Measured timing resolution between different MCP-PMTs: corrected data (blue), uncorrected data (red).}
\label{time_res}
\end{figure}
The results, see Fig.~\ref{time_res}, are obtained with the same fitting techniques exploited for $\sigma_{FEE}$. In this case the timing resolution corresponds to $\sigma_{prototype}=\sigma_{t}/\sqrt{2}$. The walk correction reduces the value of $\sigma_{prototype}$ from about 141~ps down to 103~ps improving the performance by about 25\%.\\
The Monte Carlo simulation used for the Cherenkov pattern, estimates the timing contribution of the radiator bar as $\sigma_{propagation}^{sim}\approx 63$~ps. The measured resolution $\sigma_{propagation}^{meas}$ can be obtained in first approximation by:

\begin{multline}
\sigma_{propagation}^{meas}=\sqrt{\sigma_{prototype}^{2}-\sigma_{FEE}^{2}-\sigma_{MCP-PMT}^{2}} \approx 87\mbox{~ps}\\
\label{prop}
\end{multline}

where $\sigma_{prototype}\approx103$~ps, $\sigma_{FEE}\approx40$~ps and $\sigma_{MCP-PMT}\approx37$~ps~\cite{Albert}. The difference between simulation and measurement is mainly due to synchronisation precision between the FPGAs of different TRB3 boards, and can be estimated as

\begin{equation}
\sqrt{\left(\sigma_{propagation}^{meas}\right)^{2}-\left(\sigma_{propagation}^{sim}\right)^{2}}\approx60 \mbox{~ps,}
\end{equation}

in agreement with preliminary laboratory tests. A summary of the measured timing resolutions can be found in Tab.~\ref{time_res_resume}.\\

\begin{table}[htb!]
\centering
\caption{Summary of the measured timing resolutions.}
\begin{tabular}{ccc}
\hline
Timing resolution      & not corrected & corrected \\ \hline
intrinsic $\sigma_{FEE}$ & 162~ps        & 40~ps     \\
SPE (laser studies)                    & 122~ps        & 75~ps     \\
DIRC prototype         & 141~ps        & 103~ps    \\ \hline
\end{tabular}
\label{time_res_resume}
\end{table}

The optimisation of the SPE detection was studied with different MCP-PMT gains. A direct proportionality between ToT and the gain was observed~\cite{ioRICH}. The timing resolution, as shown in Tab.~\ref{table}, is degraded by lowering the MCP-PMT gain. The ability to correct for time walk shows also a substantial deterioration at lower gain.

\begin{table}[htb!]
\centering
\caption{Timing resolution at different MCP-PMT gains.}%
\begin{tabular}{ccc}
\hline
              & \multicolumn{2}{c}{Timing Resolution} \\ \hline
Gain          & not corrected     & corrected         \\ \hline
$1\times10^5$ & 190~ps   & 180~ps   \\
$5\times10^5$ & 175~ps   & 130~ps   \\
$1\times10^6$ & 162~ps   & 40~ps    \\ \hline
\end{tabular}
\label{table}
\end{table}

\begin{figure}[htb!]
\centerline{
\includegraphics[width=0.95\columnwidth]{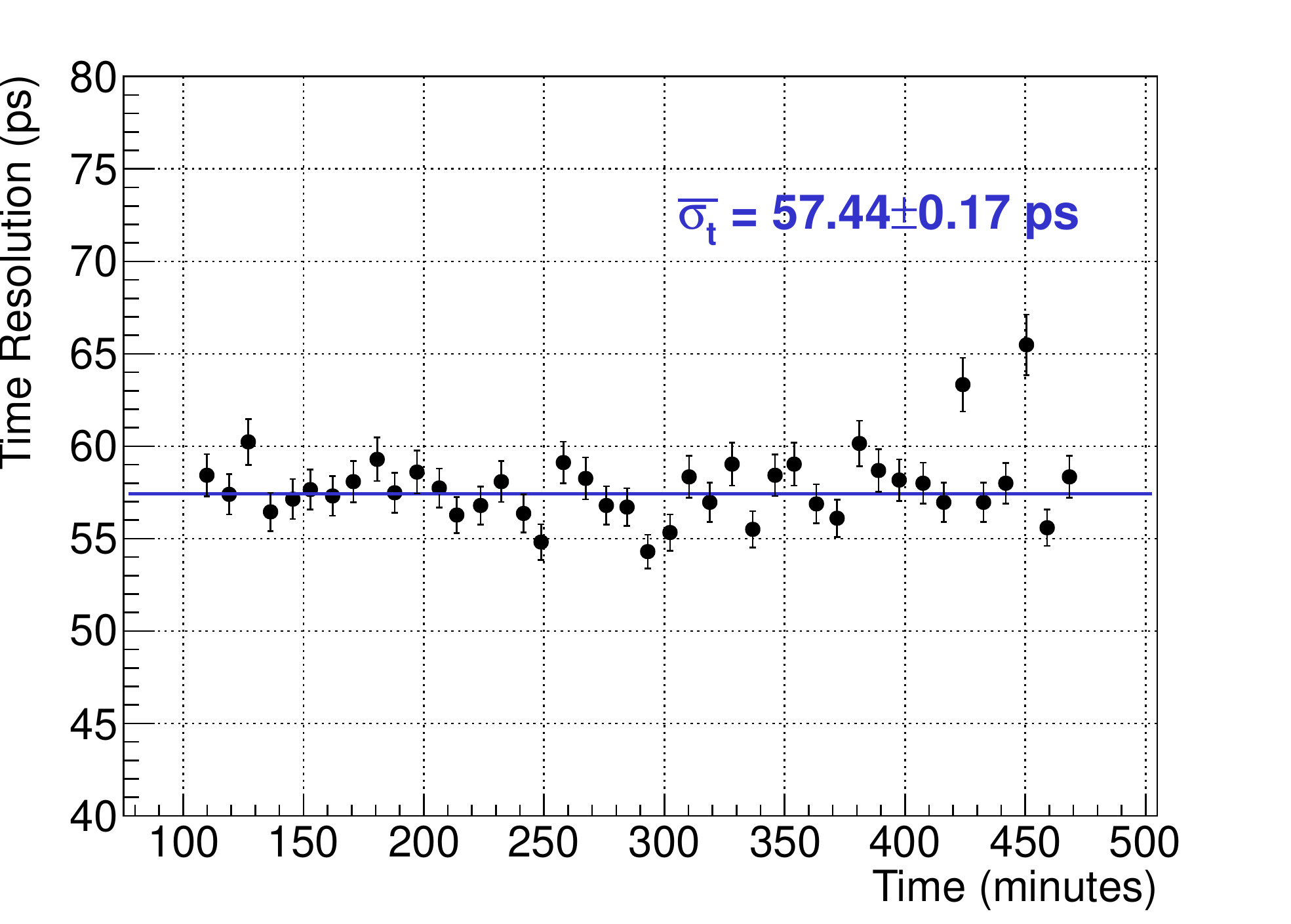}}
\caption{Measured $\sigma_{SPE}$ over an extended period of data taking.}
\label{stability}
\end{figure}

The timing performance, as established in Fig.~\ref{stability}, was stable over an extended period of time. Globally the excellent timing resolution on SPE meets the requirement of the PANDA Barrel DIRC. 

\FloatBarrier
\section{Test Experiment with a Scintillator Array}
Scintillator arrays for neutron detection are usually characterized by a large amount of channels and high background conditions. The response of the FEE to scintillator signals was investigated with a NDet prototype. The three spectrometer setup of the A1 Collaboration~\cite{A1} at MAMI offered the possibility of coincidence measurements in pion photoproduction ($e+p\rightarrow e^{'} + n + \pi^{+}$) between the $\pi^{+}$ and the particles hitting the detector.   

The prototype of the NDet is composed of two layers of plastic scintillator bars (64 bars per layer), placed atop each other and parallelly aligned. Each bar is read out by three Wavelength Shifting Fibers (WLS) which are connected via commercial optical connectors to three clear fibers (on both sides). The clear fibers are then connected to Multi-Anode PMTs\footnote{MAPMT, Hamamatsu 64~ch H7546}. The prototype was instrumented with a total of 4 MAPMT (2 per layer). The FEE cards were connected to the MAPMTs with dedicated PCB adapters, and read out with 3 TRB3 boards, for a total of 256 channels. The preamplifier Add-on cards were not used in this experiment because of the expected large signals from the scintillators. The use of discriminators for such kind of detector is justified by the high rate per channel (several~MHz). Prior experiments with a similar detector concept using an ADC-based readout had difficulties due to pile-up effects~\cite{Soeren}.

\FloatBarrier
\subsection{Time-over-Threshold response}
\begin{figure}[htb!]
\centering
\includegraphics[width=0.95\columnwidth]{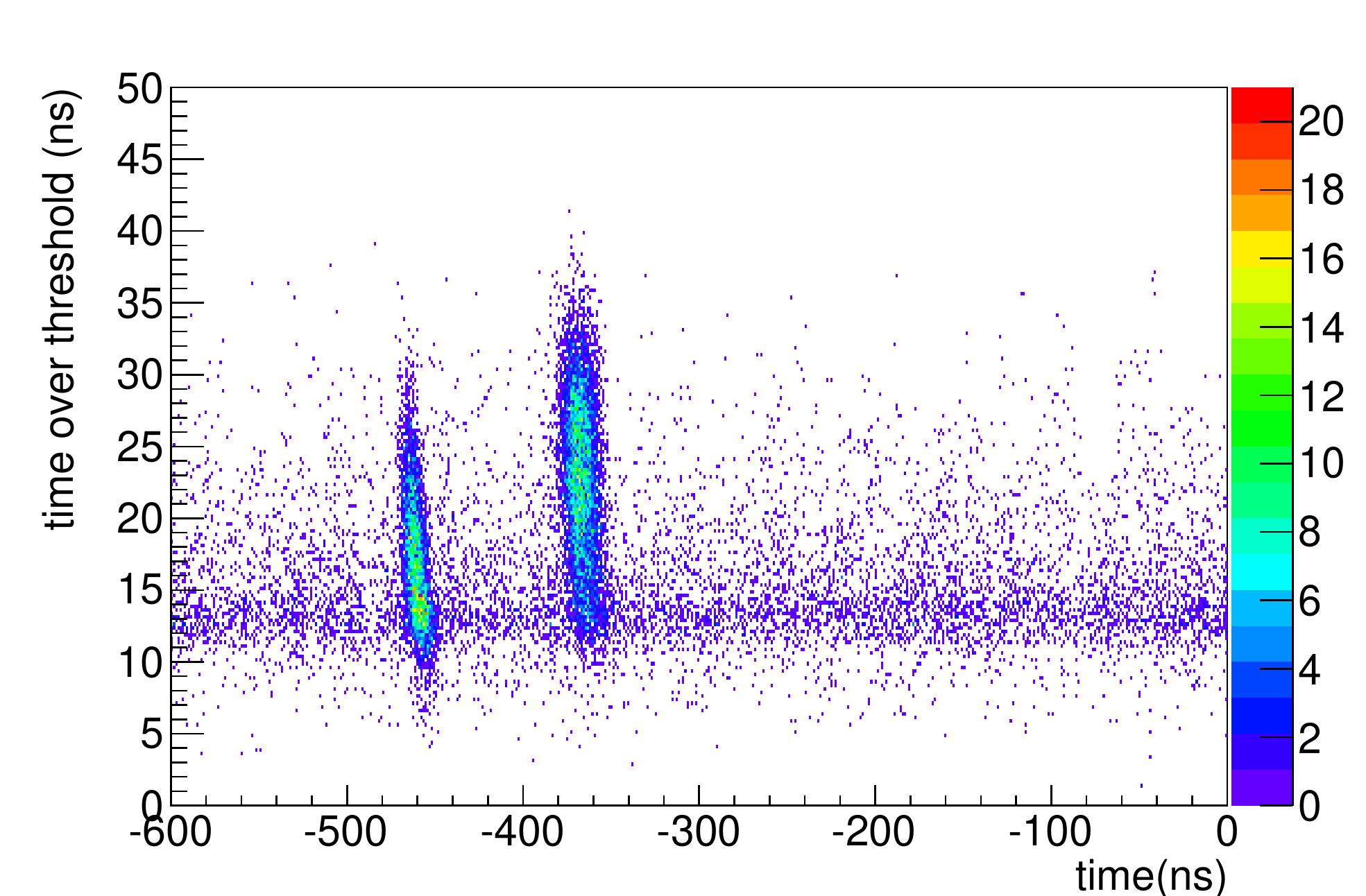}
\caption{Time-over-Threshold response as function of the relative time to the A1 trigger. The structures between [-500,-400]~ns and [-400,-300]~ns belong to MIPs and neutrons respectively, while the band with $ToT\approx12$~ns is interpreted as noise signals.}
\label{neutron_scat}
\end{figure}
 
Different particles species could be disentangled with simple cuts on the arrival time. The results in Fig.~\ref{neutron_scat} show the $ToT$ spectrum as function of the arrival time (relative to the A1 trigger at 0~ns). In the analysis a coincidence between left and right PMT channel is required, and all scintillator bars contribute to the plot. The structure between [-500,-400]~ns belongs to MIPs, which are mainly electrons from elastic scattering. The events inside [-400,-300]~ns refer to neutrons. The band with $ToT\approx12$~ns is due to noise signals. 

\begin{figure}[htb!]
\centering
\includegraphics[width=0.95\columnwidth]{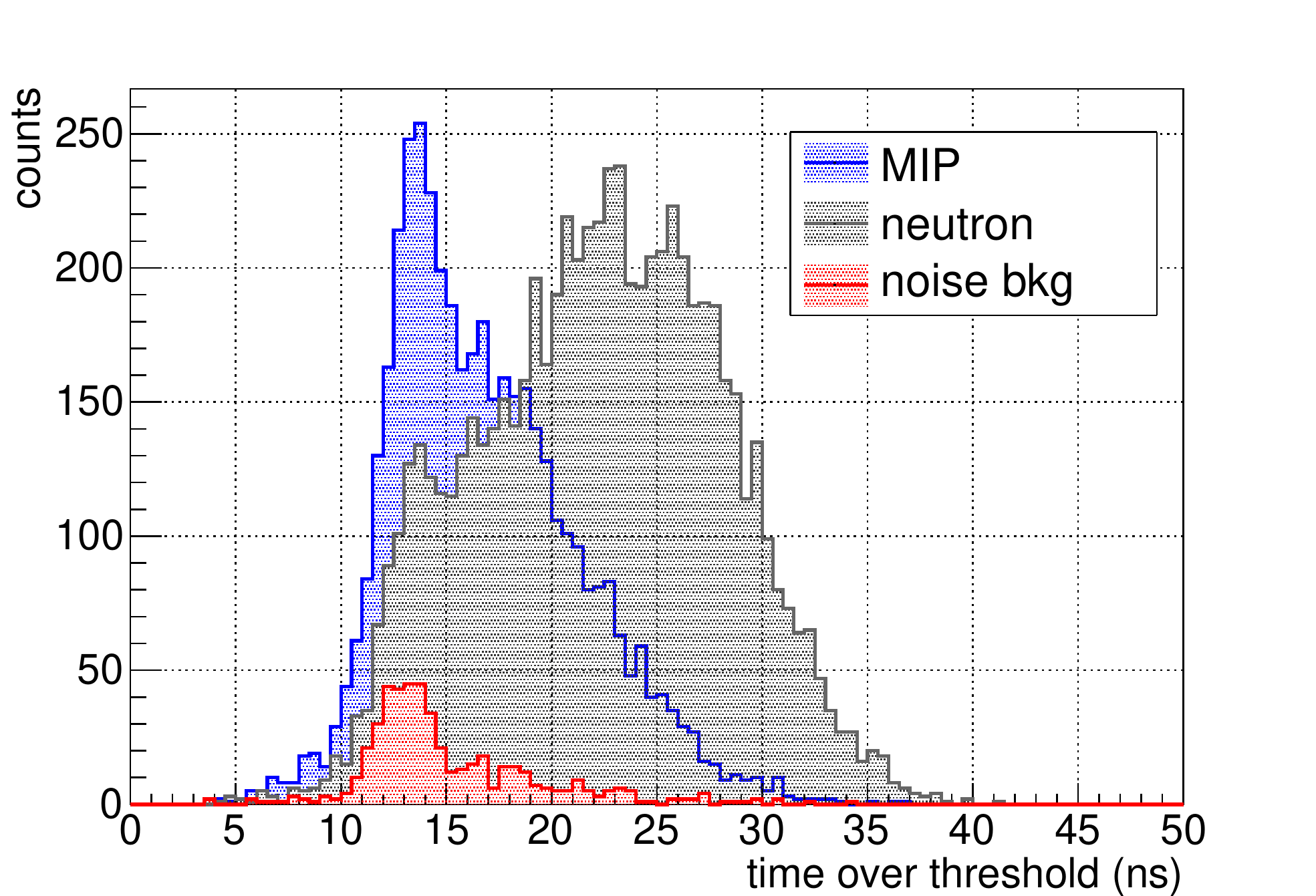}
\caption{Time-over-Threshold response for different particle species. }
\label{neutron_tot}
\end{figure}

The $ToT$ spectra for MIPs, neutrons and noise were studied selecting the corresponding region of interest. The result in Fig.~\ref{neutron_tot} shows an appreciable difference for the $ToT$. In particular the response to neutrons is characterized by a larger $ToT$ average with respect to MIPs. Normalizing both spectra and taking into account a cut around the crossing point ($\approx$20~ns), the fraction of MIPs in the neutron sample amounts to 26$\%$.   

\begin{figure}[htb!]
\centering
\includegraphics[width=0.95\columnwidth]{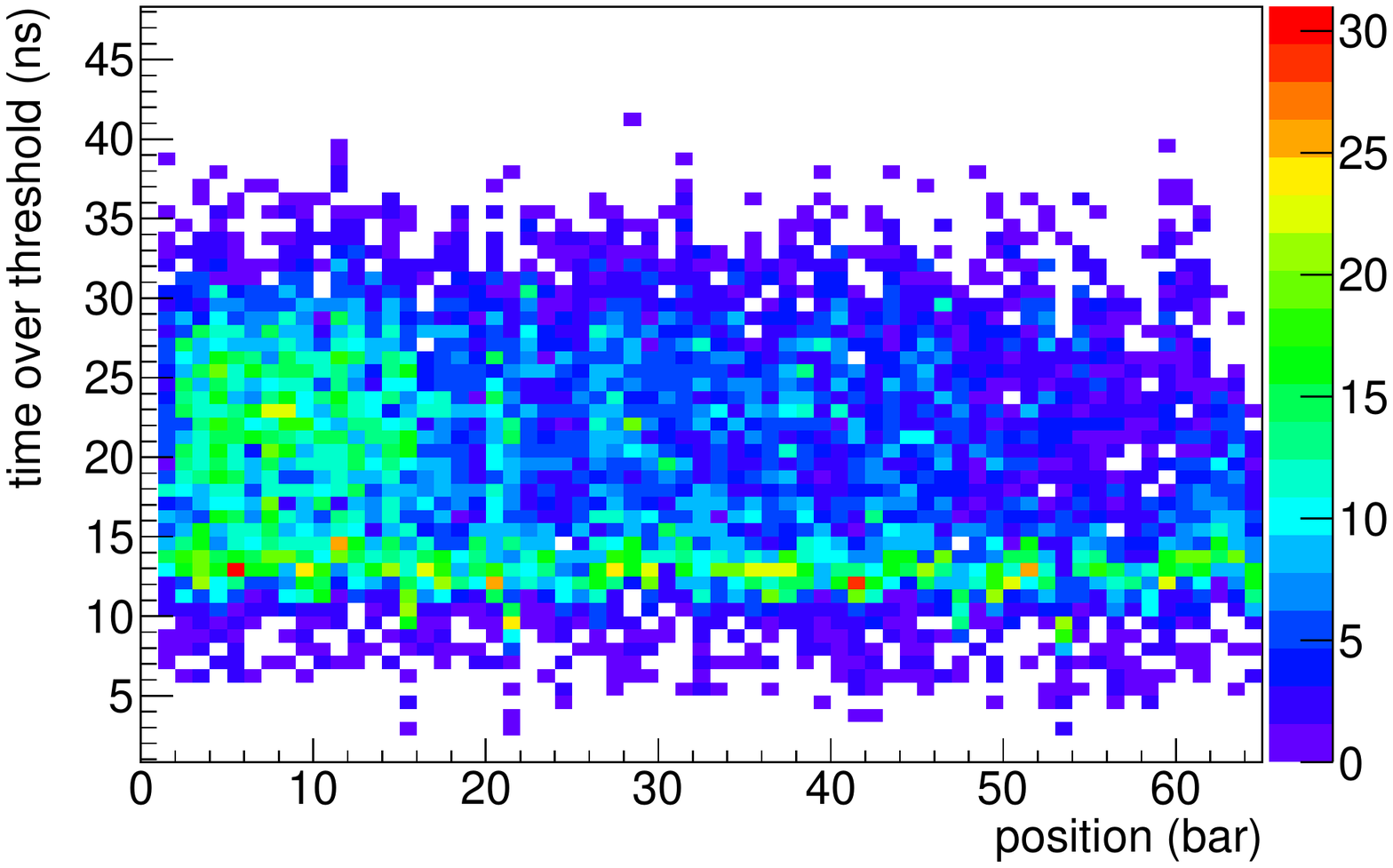}
\caption{Time-over-Threshold response across the layers of the NDet prototype. }
\label{neutron_tot_bars}
\end{figure}

The $ToT$ spatial distribution was also expected to provide additional information to disentangle neutrons and MIPs. The results in Fig.~\ref{neutron_tot_bars} show the $ToT$ spectra of incoming particles across the layers. The homogeneous band at $ToT \approx15$~ns is due to MIPs which propagate along the detectors. The neutrons cause an excess of events at higher $ToT$ (20-30~ns), which is observed between the 2$^{nd}$ and the 15$^{th}$ bar. A dedicated Geant simulation~\cite{MatthiasSim} fully agrees with the observations, and predicted also the behaviour of the first bar not showing a $ToT$ excess for neutrons.


\FloatBarrier
\subsection{Rate performance}
\begin{figure}[htb!]
\centering
\includegraphics[width=0.95\columnwidth]{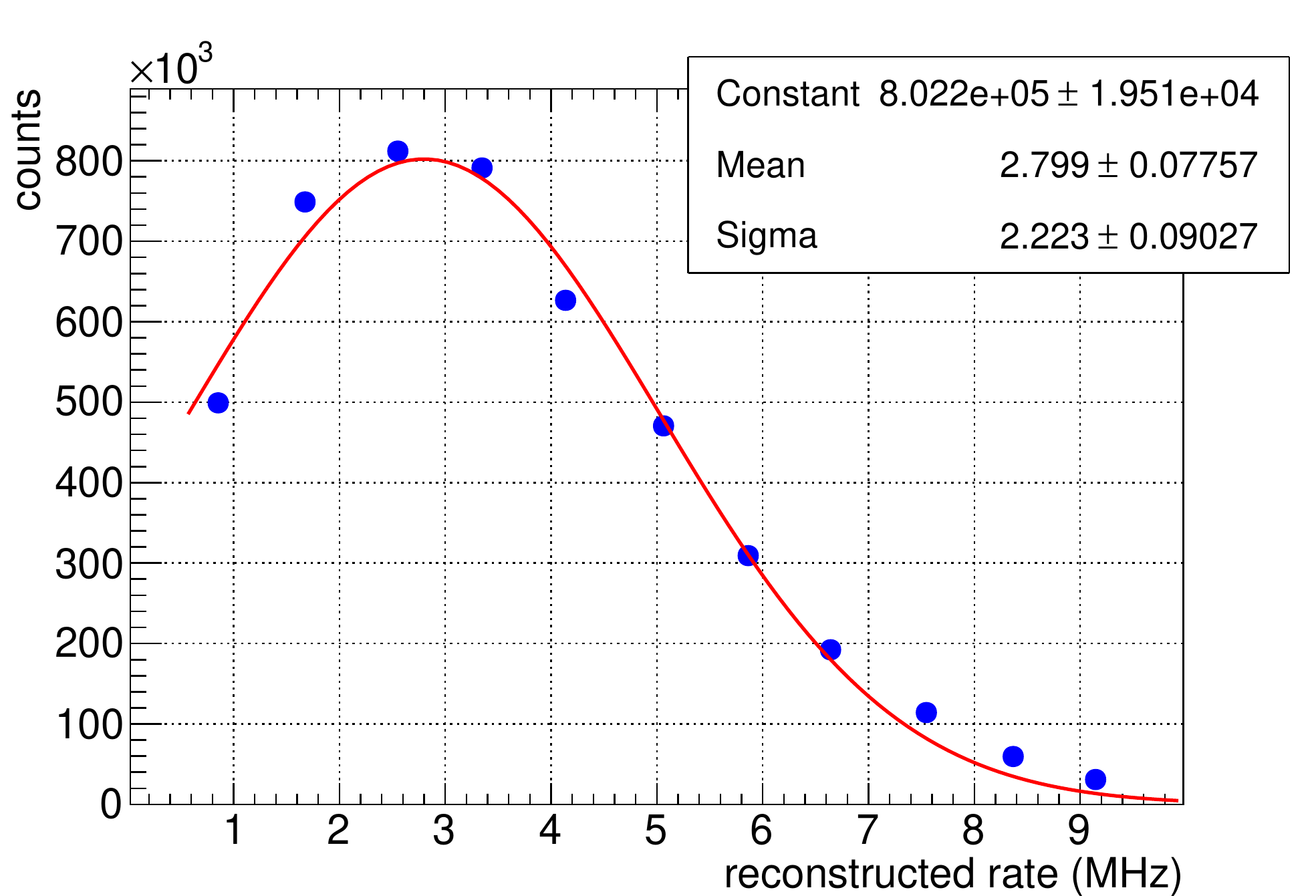}
\caption{Reconstructed rate for high beam current setting. }
\label{neutron_rate}
\end{figure}

The NDet requirements call for high count rates per channel on the order of~MHz. The NINO ASIC can sustain rates larger than 10~MHz~\cite{nino,ninocern}. However, the single ended scheme adopted for the FEE and the individual threshold had to be verified. A high beam current as used for a previous measurement of the neutron electric form factor at A1~\cite{SoerenBormio} was chosen for the studies. Considering the trigger window of 1.2~$\mu$s set for the TRB3 readout, and the number of hits per channel registered, the average rate could be reconstructed (see Fig.~\ref{neutron_rate}). A fit with a gaussian distribution gives a mean count rate of 2.8~MHz per channel with peaks up to 12~MHz. Under these conditions no saturation effects or $ToT$ modifications were discovered, which confirm the high rate capability of the developed FEE. The electronics meets the requirement of the A1 NDet .     

\FloatBarrier
\section{Conclusions}
%

The versatile design of the discriminator cards offers a compact package and low power consumption. The single ended input and individual thresholds scheme were successfully implemented and verified under realistic conditions. The developed FEE shows excellent results.\\
The response to single photon events, typical for imaging Cherenkov detectors, was successfully investigated with a fast laser pulser. The results proved the feasibility of using ToT to correct the time walk, with a timing resolution for the whole readout chain of 75~ps.  During a test experiment with a small scale DIRC prototype, the FEE achieved an excellent intrinsic resolution of 40~ps and a detector timing resolution of $\sigma_{prot}\sim100$~ps.\\ 
The results of the A1 NDet studies proved the versatility of the FEE design, which can be used to read out plastic scintillator arrays. The $ToT$ is able to deliver information about the deposited energy, fundamental for disentangling MIPs from neutrons. An additional tuning of the FEE could improve the achievable resolution, which depends on the scintillator light yield. The FEE meets the most demanding requirement of high count rates, with the possibility to reach an average of 2.8~MHz per channel without any signal degradations.\\
Further improvements will focus on the finalisation of the FEE design. In particular the availability of a 32~ch NINO chip version opens the way to higher channel density designs. The optimisation of the power consumption is another goal of future developments.

\section{Acknowledgements}
This work is partly supported by BMBF under contract no 05P12UMFP4 and the Helmholtz Graduate School for Hadron and Ion Research. The authors would like to express their gratitude to the TRB3 Collaboration for their outstanding support.


\end{document}